\documentclass[conference]{IEEEtran}
\IEEEoverridecommandlockouts
\usepackage{cite}

\usepackage{amsmath,amssymb,amsfonts}
\usepackage{algorithmic}
\usepackage{graphicx}
\usepackage{textcomp}
\usepackage{xcolor}

\usepackage{verbatim}
\usepackage{float}
\usepackage{lipsum}
\usepackage{subfigure}
\usepackage{color}
\usepackage{multirow}
\usepackage{todonotes}
\usepackage{soul}

\def\BibTeX{{\rm B\kern-.05em{\sc i\kern-.025em b}\kern-.08em
    T\kern-.1667em\lower.7ex\hbox{E}\kern-.125emX}}
\begin{document}

\title{Pyramid: Machine Learning Framework to Estimate the Optimal Timing and Resource Usage of a High-Level Synthesis Design}

\author{\IEEEauthorblockN{Hosein Mohammadi Makrani, Farnoud Farahmand, Hossein Sayadi *, Sara Bondi,\\ Sai Manoj Pudukotai Dinakarrao, Liang Zhao, Avesta Sasan, Houman Homayoun, and Setareh Rafatirad }
\IEEEauthorblockA{\textit{George Mason University, * California State University Long Beach} \\
\\
\{hmohamm8, ffarahma, sbondi, spudukot, lzhao9, asasan, hhomayou, srafatir\}@gmu.edu, * hossein.sayadi@csulb.edu}
}

\maketitle

\begin{abstract}
Field-Programmable Gate Arrays (FPGA) are powerful reconfigurable platforms that can support the efficient processing of a diverse range of applications. 
Further, the emergence of 
High-Level Synthesis (HLS) tools 
shifted the paradigm of hardware design by making the 
process of mapping high-level programming languages to hardware design such as C to VHDL/Verilog feasible. HLS tools offer a plethora of techniques to optimize designs for both area and performance, but resource usage and timing reports of HLS tools mostly deviate from the post-implementation results. In addition, to evaluate a hardware design performance, it is critical to determine the maximum achievable clock frequency. Obtaining such information using static timing analysis provided by CAD tools is difficult, due to the multitude of tool options. Moreover, a binary search to find the maximum frequency is tedious, time-consuming, and often does not obtain the optimal result. To address these challenges, we propose a framework, called Pyramid, that uses machine learning to accurately estimate the optimal performance and resource utilization of an HLS design. For this purpose, we first create a database of C-to-FPGA results from a diverse set of benchmarks. To find the achievable maximum clock frequency, we use Minerva, which is an automated hardware optimization tool. Minerva determines the close-to-optimal settings of tools, using static timing analysis and a heuristic algorithm, and targets either optimal throughput or throughput-to-area. Pyramid uses the database to train an ensemble machine learning model to map the HLS-reported features to the results of Minerva. To this end, Pyramid re-calibrates the results of HLS to bridge the accuracy gap, and enable developers to estimate the throughput or throughput-to-area of hardware design with more than 95\% accuracy and alleviates the need to perform actual implementation for estimation.
\end{abstract}

\begin{IEEEkeywords}
HLS, ensemble learning, timing estimation
\end{IEEEkeywords}

\section{Introduction}
\label{sec:intro}
The end of Dennard Scaling \cite{Dennard'74} era and the thrive to achieve high performance led to the evolution of numerous computer architecture designs. The diversity, design complexity and involved costs ceased the existence of ASICs to be the best hardware execution platforms. 

Platforms such as FPGAs emerged as the potential solution, despite the fact that FPGAs are nearly one order magnitude slower than the specialized ASICs \cite{Daniel'15, farnoud-zynq, sai-date}. FPGAs enjoy other benefits such as on-the-fly programmability, reconfigurability, and energy-efficiency \cite{hosein-asap,hosein-fccm}. Most importantly, the feasibility and  the development of hardware/software co-design tools 
facilitated computer architects or programmers to design a hardware without requiring deeper insights into hardware \cite{Holland'11, hosein-samos}. 

High-level synthesis (HLS) tools such as Xilinx's Vivado HLS~\cite{hls} and Intel's HLS \cite{intelhls} are widely used to simplify the design efforts and expedite the time-to-market. 
HLS tools translate a design written in high-level languages such as C/C++/SystemC into a low-level hardware description language (HDL). HLS shortened the learning curve of hardware accelerator design by obscuring the details of the hardware execution model. Moreover, HLS enables quick modification of a design by adding directives such as pipeline and unrolling factors that allows programmers to explore the design space. However, HLS-generated register-transfer-level (RTL) models are in general not human-readable. 

In addition, HLS tools report an estimate of the expected timing, latency, and resource utilization of the design. These reports are important as most of the time, they are the only evidence that a designer can use it to modify or optimize the design to meet the design or performance constraints. 

Throughput,  and  throughput-to-area  ratio  are  some  of  the most  important  metrics  used  for  hardware  evaluation. The maximum throughput of a design depends on the maximum clock frequency supported by the design and the way the high-level description has been synthesized. The maximum achievable clock frequency of a given HLS design can be estimated or measured at different phases of the design process. An estimation of the maximum clock frequency can be obtained from HLS timing reports \cite{choi2017hlscope+}. Despite the importance of these reports, many of them are highly inaccurate, as final resource usage and timing reports of HLS depend on the implementation phases such as logic synthesis, and place\&route, that are beyond the HLS tool capability. Therefore, it is difficult for even state-of-the-art HLS tools to accurately estimate the performance and resource utilization of a design.

On the other hand, it is also possible to calculate the maximum clock frequency in the implementation process. Timing results can be obtained after synthesis, placing\&routing, or on the real hardware design. 
The post-synthesis and post place\&route results are determined by the FPGA tools using static timing analysis. 
There are challenges associated with the static timing analysis of digital systems designs: The latest version of CAD tools provided by Xilinx (Vivado), does not have the capability to report the maximum frequency achievable for a given code. The user must request a target frequency, and the tool reports either a ``pass'' or ``fail'' for its attempt to achieve this goal. While there are 25 optimization strategies predefined in the tool, applying them sequentially, is extremely tedious and time consuming. 

To address these challenges, we propose a framework called Pyramid which uses an ensemble machine learning model to estimate the optimal throughput or throughput-to-area of a given HLS design only by using information extracted from the reports of HLS tool. To this goal, we first use Minerva tool \cite{farahmand2017minerva} -- an automated hardware optimization tool that employs a unique heuristic algorithm which is customized for frequency search using CAD toolsets. By using Minerva, we obtain results in terms of throughput, and throughput-to-area ratio for the RTL code generated by HLS tool. Then, we build a comprehensive dataset using HLS reports and corresponding Minerva results. By using this dataset, Pyramid framework trains an ensemble learning model to achieve high accuracy (more than 95\%) for the estimation tasks. In this framework, we leverage hundreds of features that can be readily extracted from the HLS reports to accurately estimate the results of Minerva without actually running the implementation flow.

\section{Challenges and Solution}
\label{sec:motiv}
\begin{figure}[t]
\centering
 \includegraphics[width=0.40\textwidth]{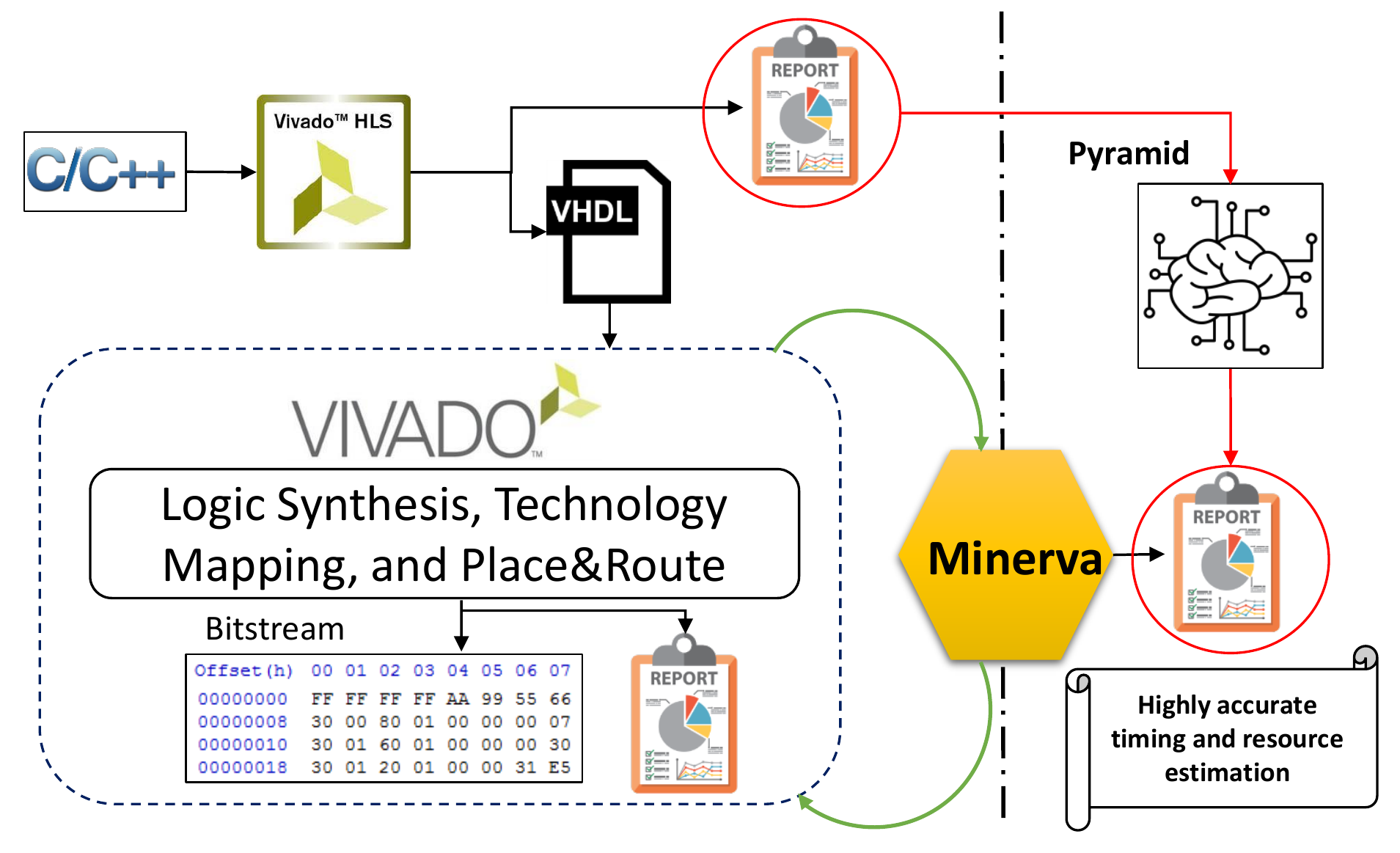}\\
\caption{HLS design flow + our approach}
\label{fig:flow}
\vspace{-1em}
\end{figure}

\subsection{Inaccuracy of the HLS report}
Figure \ref{fig:flow} (the left part) shows HLS-based design flow which starts with a high-level software program such as C, C++, or SystemC. HLS tool translates the high-level language programs into 
HDL models such as Verilog and VHDL. Additionally, HLS tools report the expected timing, and estimated resource usage.

At the HLS stage, it is hard to accurately estimate the post-HLS (implementation) results as the implementation process includes many non-trivial steps. Additionally, the reported resource utilization metrics such as the number of LUTs, flip-flops (FFs), DSPs, block RAMs (BRAMs), and timing reports is dependent on the target FPGA specification. HLS tools try to estimate resource and timing of the design by characterizing functional units and instantiated functional for each design. However, such estimation fails to capture the impact of optimization during the implementation.   

We use Root Mean Squared Error (RMSE) presented in equation \ref{eq:rm} to evaluate the accuracy of the estimation of HLS's report with respect to the results of post-implementation. We use this metric for reporting of the estimation error in this work. 

\begin{equation}
 \begin{aligned}
 \centering
 & Relative\  RMSE = \sqrt{\frac{1}{N} \sum_{n=1}^{N} (\frac{p_i - a_i}{a_i} )^2} \times 100
 \end{aligned}
  \label{eq:rm}
 \end{equation}
 
where \textit{N} is the number of samples, and \textit{$p_i$} and \textit{$a_i$} are the predicted and actual values of the sample, respectively. We want the \% relative RMSE to be as low as possible. RMSE is a standard metric in regression which is sensitive to scalability. Table \ref{tbl:hlserr} shows the Relative RMSE for timing and resource utilization for more than 90 studied benchmarks (details will be discussed later in this paper).

\begin{table}[t]
\caption{Average HLS estimation error over 90 benchmarks}
\centering
\scalebox{0.85}{
\begin{tabular}{|l|l|l|l|l|l|l|}
\hline
Devices      & \multicolumn{2}{l|}{Artix7} & \multicolumn{2}{l|}{Kintex7} & \multicolumn{2}{l|}{Virtex7} \\ \hline
Tragets      & Resource      & Timing      & Resource       & Timing      & Resource       & Timing      \\ \hline
HLS Estimate & 91.7\%        & 23.6\%      & 112.5\%        & 28.1\%      & 88.4\%         & 17.2\%      \\ \hline
\end{tabular}
}
\label{tbl:hlserr}
\end{table}

The results confirm a significant deviation 
between the HLS report and the corresponding real implementation. 
This is in fact due to the implementation process, where the HDL models go through logic synthesis, technology mapping, and placement and routing which are not considered by the HLS tools. Therefore, post-implementation reports including the actual resource usage and timing on the target FPGA significantly differs from the HLS report.

\subsection{Challenge of Hardware Evaluation}
In addition to the accurate estimate of performance metrics, evaluating them is another challenge to be addressed. 
Throughput and throughput-to-area (Freq./\#LUTs) are two most common evaluation metrics for a hardware design. The calculation of these metrics is complex, as there are several challenges in using of static timing analysis for finding the maximum clock frequency of a design.

 To show the challenge in finding maximum clock frequency, we performed synthesis and implementation for the VHDL code of a CAESAR Round 2 candidate (ICEPOLE). We generate Worst Negative Slack (WNS) results for 128 different target clock frequencies in order to observe the trend.  The target clock frequency was set to 333 MHz, and the theoretically achievable frequency (further referred to as the reference frequency) was calculated based on WNS, utilizing the following equation:
\vspace{-0.75em}
\begin{equation}
 \begin{aligned}
 \centering
Minimum\  Clock = Target\  Clock  - WNS 
 \end{aligned}
  \label{eq:wns}
 \end{equation}

\begin{figure}[!t]
\centering
 \includegraphics[width=0.45\textwidth]{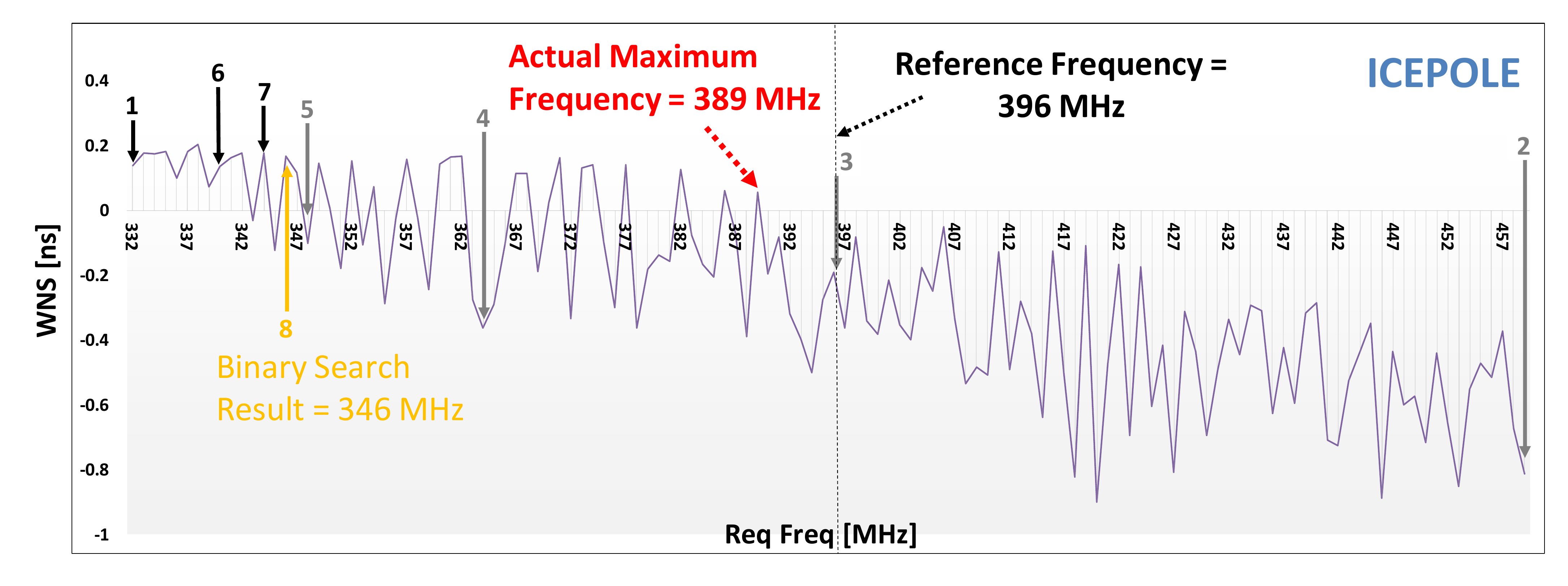}\\
\caption{Dependence of WNS on the Requested Clock Frequency and graphical representation of the binary search scheme}
\label{fig:icepole}
\vspace{-1em}
\end{figure}

In the next step, WNS results were generated for the requested
clock frequency varying from -64 to +64 MHz of the
reference frequency, with a precision of 1 MHz.  Figure \ref{fig:icepole} shows the trend. As observed in this Figure, there are fluctuations around the calculated reference clock frequency. As we can observe, the result of binary search is 346 MHz (number 8 in the figure), which is not the correct maximum frequency. Based on the ICEPOLE graph, the maximum frequency is 389 MHz. Therefore, based on the aforementioned graph, designer cannot only rely on the above equation to calculate the actual maximum clock frequency. It is important to note that these results were obtained using only default options of Vivado. Given the vast optimization strategies exist in Vivado, calculating the maximum clock frequency becomes more challenging. Another challenge is that Vivado Design Suite offers 25 predefined optimization strategies, which can be tuned to achieve a higher maximum frequency and a more optimized design. Hence, incorporating all of these strategies leads to a large exploratory space. 
Therefore, a designer cannot easily navigate all 25 optimization strategies due to the time consuming process.

\subsection{Solution}
We demonstrated that a commercial HLS tool targeting FPGAs incurs a large error of 97.5\% in estimating the resource usage. Similarly, the error for timing estimation is found to be 22.9\%. Such inaccurate estimates prevent developers from applying the appropriate set of optimization, leading to a poor trade-off. Moreover, the static timing analysis of digital systems design provided in state-of-the-art CAD tools is not able to report the maximum frequency achievable. Instead, the user must deal directly with tens of optimization strategies available in the tools which is time consuming and tedious. To circumvent such brute-force approach to find the maximum frequency, a recent work \cite{farahmand2017minerva} proposed an automated hardware optimization tool called Minerva. Minerva determines the close-to-optimal settings of tools, using static timing analysis and a heuristic algorithm developed by the authors, and targets either optimal throughput (TP) or throughput-to-area (TPA) ratio. Minerva is designed to be used to automate the task of finding optimized results. However, based on the size of the design, using Minerva may take few minutes or even several hours. Therefore, as a solution to these challenges, we propose a machine learning based framework (called Pyramid) to re-calibrate the HLS reported results and map them to the results of Minerva using ensemble machine learning method. Figure \ref{fig:flow} (the right part) shows the overview of Pyramid. In this way, without going through the time consuming process of full end-to-end implementation, using Pyramid, developers can have an accurate post-implementation estimation of resource utilization and maximum supported frequency of the design just after getting the report of their HLS design.

\section{Experimental Setup}
\label{sec:setup}
In order to build a ML model, we first create a dataset for training. 
For this purpose, we need to obtain the results of HLS tool, and its corresponding optimal implementation results. Here, we present the evaluated benchmarks, FPGA devices, and the methodology of our experiments to create the training data. 

\subsection{Benchmarks and FPGA devices}
\label{sec:appln}
To create an efficient prediction model, diversity of benchmarks plays an important role. 
Therefore, we chose a wide range of diverse benchmarks  
from Machsuit \cite{reagen2014machsuite}, S2CBench \cite{schafer2014s2cbench}, CHStone \cite{hara2008chstone}, and Rosetta \cite{zhou2018rosetta}. To further increase the diversity and the size of dataset, we included CAESAR Round 3 Candidates' HLS designs \cite{ceasar}, and a collection of 10 different image processing kernels from Xilinx xfOpenCV \cite{opencv}. We covered a total of 90 benchmarks which include a wide range of domains from simple kernels to ML and real-time video processing that reflects the
latest application trends. We categorize the benchmarks into four categories: ML, image/video processing, cryptography, and mathematical applications to validate our hypothesis that the results of the ML techniques are application dependent. We used the default version of HLS designs without applying any further directive to the designs. However, the dataset can be expanded by synthesizing designs with additional HLS optimization directives. For selecting FPGA devices, we targeted three different classes of FPGAs: low-end, medium-end, and high-end. We selected the following devices from Xilinx for evaluation: Artix7 (xc7a100tfgg484-3), Kintex7 (xc7k420tffv901-3), and Virtex7 (xc7vx980tffg1930-2). The FPGA devices are chosen based on a wide array of available resources and technologies across the spectrum of each family. The software used for our experiments are the Xilinx Vivado and Vivado HLS version 2017.2.

\subsection{Data collection}
To build a dataset, we have to extract all possible features that we can collect from HLS reports. Inputs of ML models consist of HLS features such as reource usage and timing report.
To obtain the maximum achievable clock frequency of the design, we use Minerva. Minerva is used to execute Vivado in batch mode, utilizing the Vivado batch mode Tcl scripts provided by Xilinx. An XML-based Python program is used to manage runs. This program launches Vivado with Tcl scripts that are dynamically created during run-time and later modified to perform each step of the optimization algorithm. Minerva's output includes the maximum achievable clock frequency for the design, the optimization strategy that leaded to such result, and the resource utilization for that implementation. We also parse the output of Minerva. 
Minerva's outputs will be used as the target values in our dataset. Therefore, the machine learning model will be trained to estimate resource usages for LUT, FF, DSP, and BRAM, as well as maximum clock frequency (totally 5 targets) reported by Minerva. 

As it is not possible to determine the importance of individual features in advance, we extract as many relevant features as possible from HLS reports (total 183 features). 

The flow of obtaining the results for each design is as follows: 1) we set a timing constraint (clock period) [1, 2, 4, 5, 10ns] and the FPGA device for HLS tool. 2) We run the HLS and get the VHDL files of design. We also extract all feature from HLS report. 4) We use Minerva to find the maximum clock frequency and the corresponding resource utilization for two different targets (throughput, and throughput-to-area). Minerva uses out-of-context (OOC) implementation to run Vivado. 5) We repeat the whole process for the next clock period and also for the remaining FPGA devices.

In this manner, we create a comprehensive dataset ($90 \times 5 \times 3 \times 2 = 2700$ samples) that we use for the training of ML models. We partition our dataset into two:  training/validation part, and testing part. We randomly select 20\% of our dataset for final testing as unseen data and consider the remaining 80\% as training/validation part. We perform 4-fold cross-validation on our training/validation set to train the ML models. This means that in each iteration of training, randomly 75\% of data are used for training and 25\% is used for validation. It took three months to create this dataset, using 10 servers, each of which  are  equipped with 16-core processor and 128GB memory. While this is the most time consuming part of our solution, the entire training process is only done once. We implement the ML models described in Section \ref{sec:pyramid} in Python leveraging the scikit-learn \cite{pedregosa2011scikit} and TensorFlow \cite{abadi2016tensorflow}.    

\subsection{Feature reduction}
As discussed, the obtained data includes numerous features. 
This high dimensionality may lead to computationally complex models that require longer training time, over-fitting, and are harder to interpret. Therefore, instead of accounting for all extracted features, irrelevant and redundant features are identified and removed and only a subset of features is selected. To remove redundant features, we compute the Pearson's correlation coefficient for each pair of extracted features and select one feature from each group of correlated features. To eliminate irrelevant features, we use a linear model with L2 regularization (in section \ref{subs:lr}) to fit our data. The outcome of L2-regularized linear model is a sparse estimator that eliminates (zeros) the coefficients of unimportant features. Thus, by eliminating unimportant features, we reduce the number of features from 183 to 72. Table \ref{tbl:features} shows the reduced set of non-trivial features that we eventually select from HLS reports.

\begin{table}[t]
\caption{Description of features}
\centering
\scalebox{0.9}{
\begin{tabular}{lll}
\hline 
\multicolumn{1}{l|}{Feature Category}                                             & \multicolumn{1}{l|}{Brief Description}                                                                        & \# features \\ \hline
Performance                                                               & \begin{tabular}[c]{@{}l@{}}Requested clock period, estimated \\ clock period by HLS, Uncertainty\end{tabular} & 3                \\ \hline
Resources                                                                 & \begin{tabular}[c]{@{}l@{}}Utilization and availability of\\  LUT, FF, DSP, and BRAM\end{tabular}             & 36                \\ \hline
\begin{tabular}[c]{@{}l@{}}Logic and \\ arithmetic operation\end{tabular} & Bitwidth/resource statistics of operations                                                                    & 29                \\ \hline
Memory                                                                    & Number of memory words/banks/bits;                                                                            & 2                \\ \hline
Multiplexer                                                               & Multiplexer input size/bitwidth                                                                               & 2     \\      
\hline 
\end{tabular}
}
\label{tbl:features}
\end{table}

\vspace{-1em} 
\section{Proposed Pyramid Framework}
\label{sec:pyramid}

In this section, we present the proposed Pyramid framework that employs ML models to enable fast and accurate resource and timing estimation for HLS designs. 


Use of ML has become popular in design automation, as it provides the means to accurately capture the factors impacting the accuracy of timing and resource estimation \cite{hosein-enet}.  Moreover, analytical modeling \cite{zhong2016lin-anal}, and statistic reasoning have been used to construct the estimation models as a function of multiple parameters for the evaluation of hardware design \cite{choi2017hlscope+, hlscope}. In this work, we investigate whether the models built by ML techniques are sufficiently accurate for timing and resource utilization of a design on a specific FPGA when HLS tool is used by taking into account 25 different optimization strategies available for the implementation of a design.

\subsection{Estimation models}
For the purpose of constructing timing and resource estimation models for a diverse set of benchmarks targeting different FPGA devices, we employ regression model, artificial neural network (ANN), support vector machine (SVM), and random forest (RF), where each of the predictors belongs to a different branches of ML. We explore with different models to evaluate its complexity and performance and chose the best model that can be utilized as a practical solution. 

\begin{table*}[ht]
\caption{ML estimation errors}
\centering
\scalebox{0.8}{
\begin{tabular}{|l|l|lllll|lllll|lllll|}
\hline
\multicolumn{2}{|c|}{Device}   & \multicolumn{5}{c|}{Artix7}                                                                                                             & \multicolumn{5}{c|}{Kintex7}                                                                                                            & \multicolumn{5}{c|}{Virtex7}                                                                                                            \\ \hline
\multicolumn{2}{|c|}{Resource} & \multicolumn{1}{c|}{LUT} & \multicolumn{1}{c|}{FF} & \multicolumn{1}{c|}{DSP} & \multicolumn{1}{c|}{BRAM} & \multicolumn{1}{c|}{Timing} & \multicolumn{1}{c|}{LUT} & \multicolumn{1}{c|}{FF} & \multicolumn{1}{c|}{DSP} & \multicolumn{1}{c|}{BRAM} & \multicolumn{1}{c|}{Timing} & \multicolumn{1}{c|}{LUT} & \multicolumn{1}{c|}{FF} & \multicolumn{1}{c|}{DSP} & \multicolumn{1}{c|}{BRAM} & \multicolumn{1}{c|}{Timing} \\ \hline
\multirow{4}{*}{TP}    & LR    & 27\%                     & 19\%                    & 22\%                     & 29\%                      & 18\%                        & 23\%                     & 25\%                    & 17\%                     & 21\%                      & 20\%                        & 29\%                     & 20\%                    & 21\%                     & 19\%                      & 24\%                        \\ \cline{2-2}
                       & ANN   & 12\%                     & 17\%                    & 13\%                     & 14\%                      & 16\%                        & 10\%                     & 11\%                    & 19\%                     & 16\%                      & 15\%                        & 11\%                     & 14\%                    & 18\%                     & 14\%                      & 13\%                        \\ \cline{2-2}
                       & SVM   & 22\%                     & 16\%                    & 19\%                     & 18\%                      & 19\%                        & 20\%                     & 22\%                    & 18\%                     & 15\%                      & 17\%                        & 24\%                     & 21\%                    & 16\%                     & 17\%                      & 18\%                        \\ \cline{2-2}
                       & RF    & 9\%                      & 17\%                    & 19\%                     & 14\%                      & 16\%                        & 15\%                     & 18\%                    & 12\%                     & 11\%                      & 17\%                        & 16\%                     & 20\%                    & 19\%                     & 15\%                      & 14\%                        \\ \hline
\end{tabular}
}
\label{tbl:mlerr}
\end{table*}

\subsubsection{Linear Regression} 
\label{subs:lr}
We first experimented the utilization of linear models for estimating the timing and resource information.  
In this work, we use Ridge regression model. By adding a degree of bias to the regression estimates, Ridge regression reduces the standard errors. Ridge regression solves the multicollinearity problem through shrinkage parameter $\lambda$. Also, it uses $L_{2}$ regularization method.  
We use this model to observe how much linearity exist between the features.

\subsubsection{Artificial Neural Network} 
In addition to linear modeling, we also utilize  artificial neural network (ANN) to capture the non-linearity in the data and create a non-linear model between the target and input features. 
In this work, we utilize four fully connected hidden-layer neural network with 100 neurons in each layer. However, to capture further complicated non-linear functions, one can increase the depth of neural networks. 


\subsubsection{Support Vector Machine (SVM)} 
In addition to NN to model the non-linearity, we also explore the SVM to model the data, due to the benefits of relatively lower complexity and similar performance. 
SVM analysis is a popular ML tool for nonlinear functions. SVM  is considered a non-parametric technique because it relies on kernel functions. Some problems cannot adequately be described using a linear model. In such a case, the Lagrange dual formulation in SVM allows the technique to be extended to nonlinear functions. 

\subsubsection{Random Forest} 
Random Forest is another unsupervised flexible and low-complex ML technique that can achieve better performance, even with minimal or no hyper-parameter tuning. Hence, we also analyze the performance of random forest in this work. 
One of the advantage of random forest model is that it enables to measure the relative importance of each feature on the prediction. 

To determine the best set of hyperparameters ($\lambda$ for regression and number of layers and neurons in ANN), we employed Grid search \cite{parameter-tuning}. We came up with a ANN with four hidden layers that the number of neurons in each layer is as follows: 105, 60, 44, and 30. Table \ref{tbl:mlerr} shows the error of machine learning models for resource and timing estimation. Our results show that the average errors of models built by LR, ANN, SVM, and RF are 23\%, 14\%, 19\%, and 15\%, respectively. For our estimation problem, the number of features is relatively high, and the amount of training data is limited. 
Machine learning techniques such as NN, RF, and SVM  can highly accurately represent complicated non-linear functions when a large amount of training data is provided for the model to converge. Given the available dataset, and reasonable amount of time for data collection, our experiments demonstrate that merely using general machine learning techniques fail to build accurate models. 
Therefore, this motivates us to seek ensemble learning approach to improve the accuracy of estimation. 

\subsubsection{Ensemble model} Ensemble learning is a branch of machine learning which is used to improve the accuracy and performance of general ML models by generating a set of base learners and combining their outputs for final decision \cite{sayadi2018ensemble,sayadi-2smart}. It fully exploits complementary information of different estimators to improve the decision accuracy and performance. 
In this work, we use stacked regression \cite{breiman1996stacked} method, where a number of first level estimators are combined using a second-level estimator. The key idea, is to train a second-level estimator based on the output of the first level estimators via cross-validation. It is critical to ensure that the base estimators are formed using a batch of training dataset that is different from the one used to form the new dataset. The second step is to treat the new dataset as a new problem, and employ a learning algorithm to solve it. 

To employ stacking approach, two parameters must be determined: threshold for the accuracy and the maximum number of iterations. After each stage, the accuracy must be checked. If the model meets the target estimation accuracy, then we stop the process of model creation. Otherwise, we continue the iterations until reaching to the desired accuracy or to the maximum number of iterations. When the parameters of the first-order model are determined, the accuracy of the timing and resource estimation can be evaluated. If the target accuracy is met, we cease to create sub-models. Otherwise, a higher order model is required to be further created till we reach to the threshold of iterations.

Table \ref{tbl:errtotal} shows the estimation errors of timing and utilization models created by Pyramid for two different optimization goals such as throughput (TP) and throughput-to-area (TPA) using stacking approach. Later in this section, we describe how the model is constructed. Results show that the average error of the model is only 4.7\%. Table \ref{tbl:appdep2} presents the estimation of ML techniques with regard to the benchmarks' categories. The interesting observation is that the accuracy of estimations for mathematical benchmarks are high even with linear regression. On the other hand, the resource and timing estimation of ML and cryptography benchmarks are lower with general ML techniques. However, ensemble learning shows a good accuracy for all benchmarks.      

Taking into account the size of our dataset and high dimensional features, solely using a general machine learning technique incurs a large error. To overcome this, the proposed Pyramid framework helps to create an accurate model for estimating the optimal throughput and throughput-to-area of a HLS design. 

\begin{table}[t]
\caption{Average error of Pyramid estimations}
\centering
\scalebox{0.8}{
\begin{tabular}{|c|c|c|c|c|c|c|}
\hline
\textbf{Devices}     & \multicolumn{2}{c|}{\textbf{Artix7}} & \multicolumn{2}{c|}{\textbf{Kintex7}} & \multicolumn{2}{c|}{\textbf{Virtex7}} \\ \hline
\textbf{Tragets}     & \textbf{Resource}  & \textbf{Timing} & \textbf{Resource}  & \textbf{Timing}  & \textbf{Resource}  & \textbf{Timing}  \\ \hline
\textbf{Pyramid-TP}  & 6.3\%              & 3.8\%           & 5.2\%              & 4.1\%            & 4.9\%              & 4.4\%            \\ \hline
\textbf{Pyramid-TPA} & 4.8\%              & 3.5\%           & 4.7\%              & 4.6\%            & 4.8\%              & 4.9\%            \\ \hline
\end{tabular}
}
\label{tbl:errtotal}
\end{table}

One of the issues in using stacking approach is determining the correct weights to combine the models, as topology and hyper-parameters of the model are the crucial for the performance of ensemble learning. To address this challenge and alleviate the the burden on the end-user, we suggest the following guidelines to be followed by developers to facilitate adopting our framework and makes it possible to reproduce the results presented here.


\subsection{Guidelines}
\begin{figure}[t]
\centering
 \includegraphics[width=0.35\textwidth]{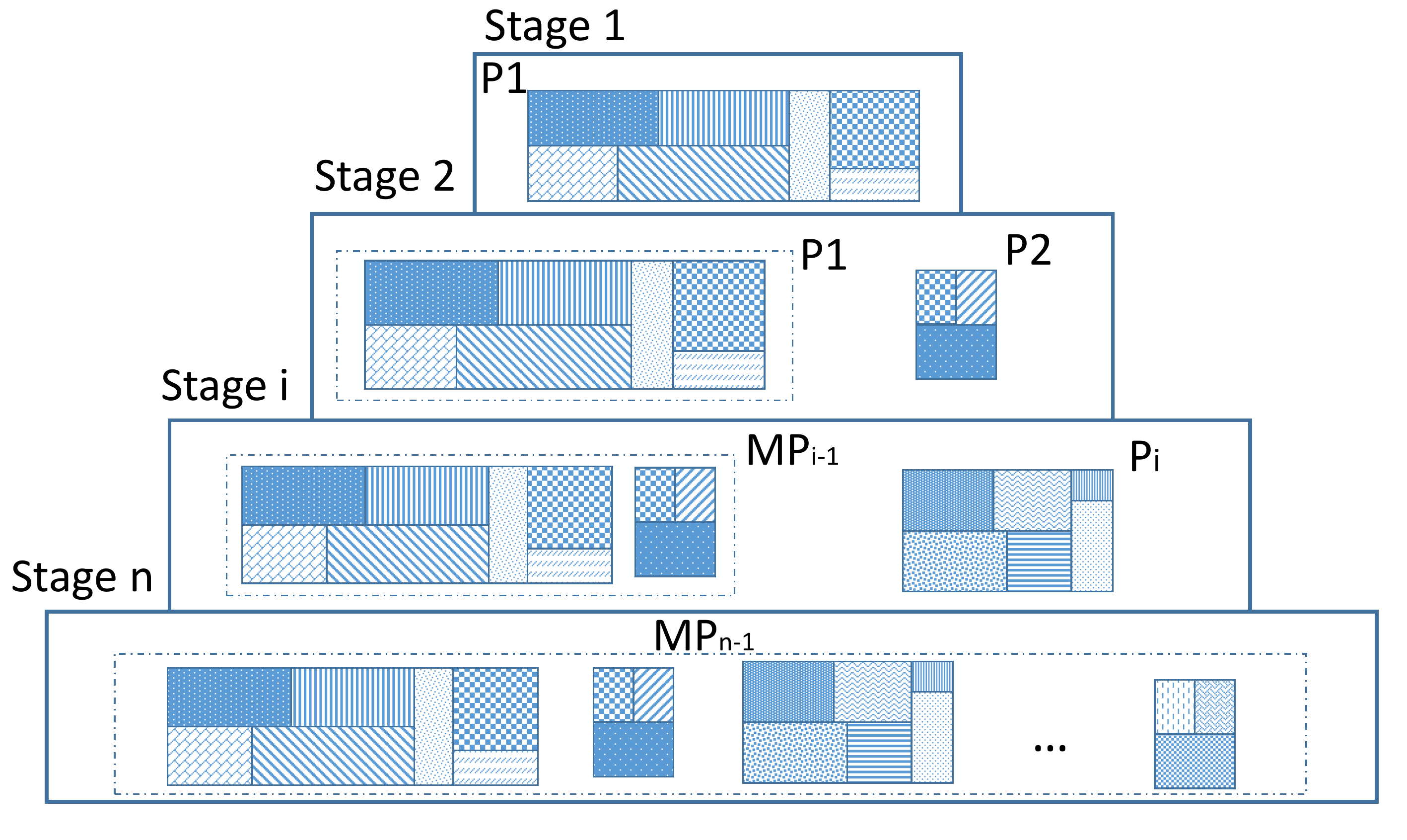}\\
\caption{Overview of stacking approach}
\label{fig:pyramid}
\vspace{-1em}
\end{figure}

Figure \ref{fig:pyramid} shows the overview of stacking approach. In each stage, sub-model can be created by any arbitrary machine learning techniques such as LR or NN and they can be considered as a black-box.  As an example, we employed a simple three-layer fully connected neural network with 20 hidden neurons employed to create the sub-models. We set the target accuracy and the threshold for the maximum number of iteration to 99\% and 50 respectively. In the first stage, a single sub-model (P1) is created as a function of the features from Bootstrap samples of main dataset. Each time, we randomly select 20\% of samples with replacement. Bootstrapping is important here as the dataset is relatively small. 
For the second stage, we employ another neural network (P2) to model the variation in the estimation of designs that is not modeled by P1. The boxes of different shades in Figure \ref{fig:pyramid} signify the parts of dataset which have been modeled. 

Now, we can build a primary mixed model, MP1, through the combination of our first two sub-models P1 and P2 as follow: $MP1 = \alpha1P1 + \alpha2P2$. P1 and P2 represent the estimated timing by sub-models, and $\alpha$s are the coefficients corresponded to the learning rate. To simplify the procedure, we set the $\alpha$ values to 0.1. We repeatedly create sub-models and add them to the mixed model. 
If the model achieves the desired accuracy before the total number of iteration, the final model is obtained. This is called the first-order model (FM1). If we don not reach to the target accuracy after the threshold on the number of iterations, above procedure can be repeated to create another mixed model (FM2) or the user can end the process. In our case, we didn't meet our target accuracy after 50 iterations but we reached to more than 95\% accuracy which is acceptable for our case and we stopped the training procedure.

It is noteworthy to consider the following rules: A lower value for the learning rate increases the number of sub-models required to create the first-order model at a specific accuracy. On top of that, changing the parameter of each sub-model affects the total number of sub-models to create the first-order model. Moreover, increasing the complexity of sub-models results in fewer sub-models required to achieve the maximum accuracy, at a given learning rate.

\begin{table}[t]
\caption{Average error of ML techniques for different benchmarks}
\centering
\scalebox{0.9}{
\begin{tabular}{|c|c|c|c|c|c|c|c|c|}
\hline
\textbf{\begin{tabular}[c]{@{}c@{}}Benchmark's\\ category\end{tabular}} & \multicolumn{2}{c|}{\textbf{\begin{tabular}[c]{@{}c@{}}Machine\\ Learning\end{tabular}}} & \multicolumn{2}{c|}{\textbf{\begin{tabular}[c]{@{}c@{}}Img/Vid\\ Processing\end{tabular}}} & \multicolumn{2}{c|}{\textbf{Crypto.}} & \multicolumn{2}{c|}{\textbf{Mathe.}} \\ \hline
\textbf{Targets}                                                   & \textit{Res}                                & \textit{Tim}                               & \textit{Res}                                 & \textit{Tim}                                & \textit{Res}      & \textit{Tim}      & \textit{Res}      & \textit{Tim}     \\ \hline
\textbf{LR}                                                        & 29\%                                        & 25\%                                       & 17\%                                         & 16\%                                        & 38\%              & 22\%              & 11\%              & 8\%              \\ \hline
\textbf{ANN}                                                       & 17\%                                        & 14\%                                       & 13\%                                         & 11\%                                        & 19\%              & 14\%              & 8\%               & 7\%              \\ \hline
\textbf{SVM}                                                       & 22\%                                        & 19\%                                       & 18\%                                         & 17\%                                        & 23\%              & 18\%              & 10\%              & 7\%              \\ \hline
\textbf{RF}                                                        & 16\%                                        & 16\%                                       & 14\%                                         & 12\%                                        & 20\%              & 15\%              & 9\%               & 7\%              \\ \hline
\textbf{Ensemble}                                                  & 6\%                                         & 5\%                                        & 4\%                                          & 3\%                                         & 5\%               & 4\%               & 4\%               & 3\%              \\ \hline
\end{tabular}
}
\label{tbl:appdep2}
\end{table}
\section{Related Work}
\label{sec:relat}
Here, we present and compare the proposed Pyramid framework with some of the recent works whose goals are similar. 
Use of machine learning has become popular in design automation \cite{dai2018fast,makrani2019xppe, new-dac, new2, n1, n2}. 
The difference of our work with Dai et al. is as follow: they mapped the report of HLS tool to the correspondent result of the implementation. However, we explained that the latest version of Vivado tool, does not have the capability to report the maximum frequency achievable for the corresponding design  and the calculated maximum frequency based on WNS reported by the tool is wrong. We addressed this issue by leveraging Minerva in our work. Makrani et. al. used ML to estimate the speed up of  an hardware accelerator on arbitrary FPGA over an ARM processor. 
In \cite{Daniel'15}, the performance prediction for Zynq-SoC is proposed which estimates the performance based on the execution 
time of an application on the FPGA.  In \cite{kapre2015intime}, the authors present InTime, a machine learning approach, supported by a cloud-based compilation infrastructure,
to automate the selection of FPGA CAD tool parameters and minimize the TNS (total negative slack) of the design. InTime does not have the capability to find the actual maximum frequency with positive TNS near zero. 
Recently, several studies applied machine learning for auto-tuning of frameworks to explore design space of tool parameters for improving FPGA synthesis and implementation \cite{13qor, 14qor, hlspredict, shervin, lamda}. Machine learning was used in HLS to reduce the number of design candidates required to run for implementation \cite{15qor}. In \cite{16qor}, a three-layer ANN model was trained to estimate the resource usage of post-implementation from pre-characterized area models of a small set of template-based designs. Different from prior works, our work uses machine learning to re-calibrate the results of HLS reports to provide an accurate post-implementation estimate of resource utilization and maximum supported frequency of HLS design for developers.  

\section{Conclusions}
\label{sec:conclu}
In this work, we thoroughly studied the main challenges of evaluating an HLS design: the inaccuracy of HLS reports in timing and resource utilization. Moreover, HLS reports lack insights for finding the optimal throughput or throughput-to-area of the generated RTL design. To address these challenges, we proposed Pyramid, a framework that uses ensemble learning technique to bridge the accuracy gap between HLS report and the optimal achievable throughput or throughput-to-area of the HLS design. To achieve this, we first used Minerva, an automated hardware optimization tool, to find the maximum clock frequency and resource utilization of the RTL code of the design generated by HLS tool. We then, used stacking approach to map the features extracted from HLS reports to Minerva's output. 
As the dimensionality of features is high and the sample size of the dataset is not significantly large, the obtained accuracy of several studied ML estimators is found to be low. In response, the model created by Pyramid framework using ensemble learning is shown to have more than 95\% accuracy. 

\bibliographystyle{IEEEtran}
\bibliography{litreture}
\end{document}